\documentclass[epsfig,graphics,onecolumn,showpacs,floatfix,mathbbm,prl,superscriptaddress]{revtex4-1}

\usepackage[utf8x]{inputenc}
\usepackage{hyperref}
\hypersetup{
  colorlinks   = true, 
  urlcolor     = blue, 
  linkcolor    = blue, 
  citecolor    = red 
}
\usepackage{graphicx}
\usepackage{amsmath}
\usepackage{amsfonts}
\usepackage{amssymb}
\usepackage{float}
 
\newcommand{\im}{\mathrm{i}}

\newcommand{\unit}[1]{\,\mbox{#1}}
\newcommand{\ket}[1]{\left| #1\right\rangle}
\newcommand{\bra}[1]{\left\langle #1 \right|}

\newcommand{\abs}[1]{\left| #1 \right|}
\newcommand*{\bigchi}{\mbox{\Large$\chi$}}
\newcommand{\ba}{\begin{eqnarray}}
\newcommand{\ea}{\end{eqnarray}}

\newcommand{\kW}{\bold{k}_{\rm W}}
\newcommand{\kw}{\bold{k}_{\rm w}}
\newcommand{\kR}{\bold{k}_{\rm R}}
\newcommand{\kr}{\bold{k}_{\rm r}}
\newcommand{\kc}{\bold{k}_{\rm c}}
\newcommand{\kp}{\bold{k}_{\rm p}}

\begin{document}

\title{Storing single photons emitted by a quantum memory on a highly excited Rydberg state}
\author{Emanuele Distante\footnote{These authors contributed equally to this work.}}

\email{emanuele.distante@icfo.es}
\affiliation{ICFO-Institut de Ciencies Fotoniques, The Barcelona Institute of Science and Technology, 08860 Castelldefels (Barcelona), Spain}
\author{Pau Farrera$^{\rm a}$}
\affiliation{ICFO-Institut de Ciencies Fotoniques, The Barcelona Institute of Science and Technology, 08860 Castelldefels (Barcelona), Spain}

\author{Auxiliadora Padr\'{o}n-Brito}
\affiliation{ICFO-Institut de Ciencies Fotoniques, The Barcelona Institute of Science and Technology, 08860 Castelldefels (Barcelona), Spain}
\author{David Paredes-Barato}
\affiliation{ICFO-Institut de Ciencies Fotoniques, The Barcelona Institute of Science and Technology, 08860 Castelldefels (Barcelona), Spain}
\author{Georg Heinze}
\affiliation{ICFO-Institut de Ciencies Fotoniques, The Barcelona Institute of Science and Technology, 08860 Castelldefels (Barcelona), Spain}
\author{Hugues de Riedmatten}
\email{hugues.deriedmatten@icfo.es}
\affiliation{ICFO-Institut de Ciencies Fotoniques, The Barcelona Institute of Science and Technology, 08860 Castelldefels (Barcelona), Spain}
\affiliation{ICREA-Instituci\'{o} Catalana de Recerca i Estudis Avançats, 08015 Barcelona, Spain}

\date{\today}
\maketitle

\section{Abstract}
\textbf{Strong interaction between two single photons is a long standing and important goal in quantum photonics. This would enable a new regime of nonlinear optics and unlock several applications in quantum information science, including photonic quantum gates and deterministic Bell-state measurements. In the context of quantum networks, it would be important to achieve interactions between single photons from independent photon pairs storable in quantum memories. So far, most experiments showing nonlinearities at the single-photon level have used weak classical input light. Here, we demonstrate the storage and retrieval of a paired single photon emitted by an ensemble quantum memory in a strongly nonlinear medium based on highly excited Rydberg atoms. We show that nonclassical correlations between the two photons persist after retrieval from the Rydberg ensemble. Our result is an important step towards deterministic photon-photon interactions, and may enable deterministic Bell-state measurements with multimode quantum memories.}

\section{Introduction}
Efficient photon-photon interactions require a highly nonlinear medium which strongly couples with a light field, a single-photon source compatible with the medium and the ability to coherently map the photon in and out of the nonlinear medium  \citep{Chang2014}. In addition, for quantum repeaters applications for long distance quantum communication, the single photon should be part of a correlated photon pair generated by a quantum memory (QM) which allows for synchronization along the communication line \cite{Sangouard2011}.
Nonlinearity at the single-photon level has been demonstrated with a variety of systems, including single atoms and atomic ensembles \cite{Dayan2008,Reiserer2013,Reiserer2014,Tiecke2014,Shomroni2014,Chen2013,Fushman2008,Volz2012,Ritter2012,Piro2010,Hacker2016} as well as nonlinear crystals albeit with small efficiency \cite{Guerreiro2014a}.
However, the coupling of true single photons with a highly nonlinear medium has been demonstrated so far only with single atoms \cite{Ritter2012,Piro2010}. These systems are inherently nonlinear but suffer from low light-matter coupling in free-space and therefore require experimentally challenging high-finesse cavities. 

Using highly excited Rydberg states of atomic ensembles can be a simpler alternative. The atomic ensemble ensures a strong light-matter coupling and the dipole-dipole (DD) interactions between Rydberg states enable strong, tunable nonlinearities. For a sufficiently dense ensemble ($\rho\sim 10^{12} \, \mathrm{cm}^{-3}$) and at sufficiently high quantum number of the Rydberg state ($n \geq 60$), nonlinear response at the single-photon level has been already demonstrated \cite{Peyronel2012,Dudin2012,Dudin2012a,Firstenberg2013,Maxwell2013} and has been exploited to implement a number of operations with weak coherent states (WCSs) \cite{Tiarks2014a,Baur2014a,Gorniaczyk2014a,Tiarks16}. Entanglement between a light field and a highly excited Rydberg state \cite{Li2013} has also been recently demonstrated.

While single-photon nonlinearities have been demonstrated with weak coherent states, efficient quantum information processing using this system require two additional steps. First, a single-photon source that matches the frequency and the sub-MHz spectral bandwidth of the Rydberg excitation, and, second, the ability to store and retrieve the input single photon. The latter is of key importance for implementing high-fidelity photonic quantum operations using excited Rydberg states \cite{Paredes-Barato2014,Khazali2015,Gorniaczyk2014a}, and, in addition, it has been shown to be beneficial to enhance the nonlinear response of this kind of systems \cite{Distante2016}. While storage and retrieval of a single photon transmitted between remote atomic ensembles has been achieved in ground states \cite{Chaneliere2005,Eisaman2005,Choi2008,Lettner2011,Zhou2012} or low-lying Rydberg states \cite{Ding2015}, storing it in a highly nonlinear Rydberg ensemble presents additional experimental challenges, such as high sensitivity to stray fields, stronger motional-induced dephasing due to the large wavelength mismatch between the single photon and the coupling laser, weak transition oscillator strength requiring higher intensity of the coupling beam, as well as strong focusing of the single-photon needed to achieve non-linearity at low light power. These challenges make it more difficult to achieve the required signal-to-noise ratio to preserve the quantum character of the stored and retrieved field. 

Here we demonstrate storage and retrieval of a paired and synchronizable single photon in a highly nonlinear medium based on excited Rydberg atomic states of a cold atomic ensemble. {This is realized  by using a photon source  based on a read-only cold atomic ensemble quantum memory \cite{Duan2001c} with which we can generate pairs of non-classically correlated photons that fulfil the frequency and the narrow bandwidth requirement of the Rydberg medium. In the generation stage (site A in Figure \ref{fig:fig1}), after a successful heralding event a single photon is emitted at a programmable delay time $t_{\rm A}$ allowing for potential synchronization between different pair sources. The photon is then collected into an optical fibre and sent to a remote atomic ensemble (site B in Figure \ref{fig:fig1}) where it is stored as a collective Rydberg excitation and retrieved after a storage time $t_\mathrm{B}$
 The storage and retrieval in high lying Rydberg states is realized with sufficiently high signal-to-noise ratio (SNR$>$20) to enable the demonstration of  highly non-classical correlations between the heralding photon  and the highly excited Rydberg collective excitation, and preservation of the single photon character of the retrieved field.} Finally, we also demonstrate the highly nonlinear response of our medium with weak coherent states containing tens of photons. The last result is obtained in a cloud with moderate density ($\rho \sim 10^{10} \, \rm cm^{-3}$) and can be easily improved to reach single-photon nonlinearity via upgraded well-known atom trapping techniques. Combining a source of narrow-band correlated single photons with a highly nonlinear medium at high signal-to-noise ratio, our system is a building block for future quantum networks with deterministic operations.

\section{Results}
\subsection{Experimental set-up}

\begin{figure}[h]
 \includegraphics[width=\linewidth]{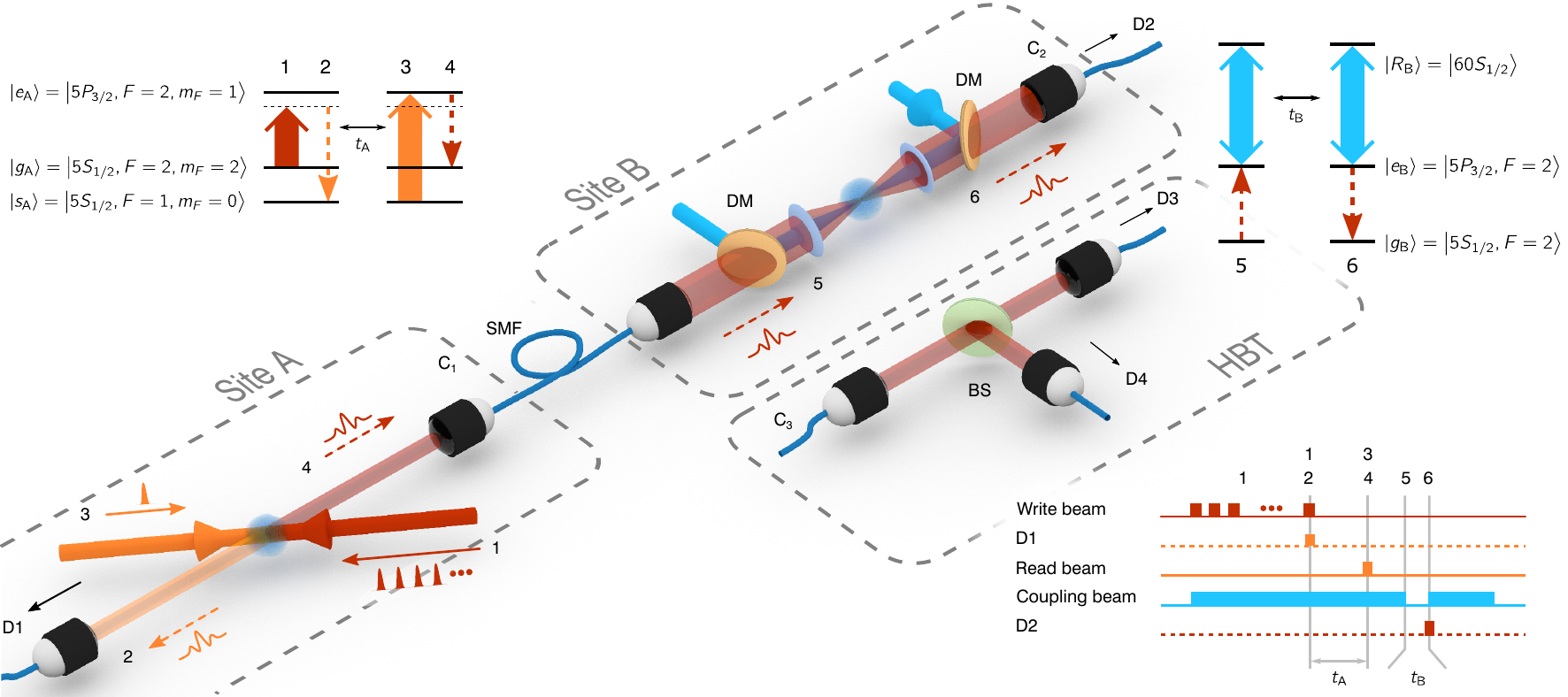}
 \caption{\textbf{Experimental setup with relevant atomic transitions and pulse sequence}. Following the numbering in the pulse sequence, in site A we (1) send a series of write pulses (red solid arrow), (2) probabilistically detect a write photon (orange dashed line) by single-photon detector (SPD)) D1, (3) send a an intense read pulse (orange solid arrow) after a storage time $t_{\rm A}$  generating  deterministically (4) a read photon (red dashed line) which is sent to Site B through a single mode fiber (SMF). In site B, a counter propagating, coupling beam (blue arrow) converts the read photon into a slowly propagating dark-state polariton. Here we (5) switch off the coupling beam, storing the read photon, and (6) switch it on again after a storage time $t_{\rm B}$ retrieving the photon which is detected by SPD D2.  The coupling beam and the read photon are both focused by the same pair of aspheric lenses and are combined and separated by dichroic mirrors (DM). Hanbury Brown-Twiss (HBT) setup is shown in another box. The field to be analyzed emerges from the SMF at position C$_3$, it is split by a 50:50 beam splitter (BS) and detected by two detectors, D3 and D4 afterwards. To analyze the photon statistic before and after storage in the Rydberg state, we connect the HBT setup either at position C$_1$ or at C$_2$.} 
\label{fig:fig1}
\end{figure}

A schematic of the experiment is shown in Figure \ref{fig:fig1}. In site A, we implement a photon-pair source with controllable delay, using a cold atomic quantum memory based on the Duan-Lukin-Cirac-Zoller (DLCZ) scheme \cite{Duan2001c,Kuzmich2003}. We use a cold atomic ensemble of $^{87} \rm Rb$ atoms. Atoms initially prepared in the ground state $\ket{g_{\rm A}} = \ket{5 S_{1/2}, F=2,m_{F}=2}$ are illuminated with a series of weak coherent pulses at $780 \, \rm nm$ (\emph{write pulses}) red detuned by $\Delta=40 \, \rm{MHz}$ with respect to the $\ket{g_{\rm A}}\rightarrow\ket{e_{\rm A}}= \ket{5{\rm} P_{3/2}, F=2,m_{F}=1}$ transition so that a \emph{write photon} is probabilistically created via Raman scattering and detected on the single-photon detector (SPD) D1. This heralds a single collective excitation in the state $\ket{s_{\rm A}} = \ket{5{\rm} S_{1/2}, F=1,m_\mathrm{F}=0}$ (see Supplementary Note 1). The excitation can be deterministically read out after a controllable storage time $t_\mathrm{A}$ by means of a strong, counterpropagating read pulse on resonance with the $\ket{s_{\rm A}}\rightarrow\ket{e_{\rm A}}$ transition. The read pulse creates a 350 ns long (FWHM) \textit{read photon} in a well defined spatio-temporal mode resonant with the $\ket{g_{\rm A}}\rightarrow\ket{e_{\rm A}}$ transition. The read photon is collected and sent through a $10 \,\rm m$ single-mode fiber (SMF) to site B, where a separate ensemble of cold  $^{87} \rm Rb$ atoms is prepared in the state $\ket{g_{\rm B}}=\ket{5{\rm} S_{1/2}, F=2}$.{We estimate that the probability to obtain a single photon in front of ensemble B conditioned on the detection of a write photon (the heralding efficiency) is $\eta_H=0.27$. At site B}, a coupling beam at $480 \, \rm nm$ resonant with the $\ket{e_{\rm B}}\rightarrow\ket{R_{\rm B}}$, transition creates the condition for electromagnetically induced transparency (EIT) \cite{Fleischhauer2000,Fleischhauer2002,Fleischhauer2005} (see Supplementary Note 3 and Supplementary Figure 2), where $\ket{e_{\rm B}} = \ket{5{\rm} P_{3/2}, F=2}$ and $\ket{R_{\rm B}} = \ket{60{\rm} S_{1/2}}$. This converts the read photon into a slow-propagating Rydberg \emph{dark-state polariton} (DSP, see Supplementary Note 4). By adiabatically switching off the coupling beam, the read photon is stored as single collective Rydberg atomic excitation \cite{Liu2001} and the state of the ensemble reads: 
\begin{equation}\label{CollectiveRydberg}
\ket{\psi_{\rm B}} = \frac{1}{\sqrt{N_{\rm B}}}\sum_{j=1}^{N_{\rm B}} e^{-i (\kp+\kc) \bold{r}_j}\ket{g_{{\rm B}_1} ... R_{{\rm B}_j} ... g_{{\rm B}_{N_{\rm B}}}},
\end{equation}
where $N_{\rm B}$ is the number of atoms in the interaction region and $\kp$ and $\kc$ the wavevector of the single photon and of the coupling beam respectively. The stored excitation is retrieved after a storage time $t_{\rm B}$ by switching the coupling beam back on and detected by a SPD D2 (see Supplementary Figure 3). The read photon waveform in ensemble A can be tailored by shaping the read pulse \cite{Farrera2016} in order to maximize the signal-to-noise ratio of the storage in site B. 
{Notice that to match the frequency of the single photon emitted at site A, we have to employ at site B the $\ket{5{\rm} P_{3/2}, F=2}\rightarrow\ket{n{\rm} S_{1/2}}$ instead of the most commonly used and stronger $\ket{5{\rm} P_{3/2}, F=3}\rightarrow\ket{n{\rm} S_{1/2}}$ transition  \cite{Gorniaczyk2014a,Tiarks16,Maxwell2013,Firstenberg2013,Peyronel2012}. This makes it more challenging to reach high storage efficency of the single photon into the collective Rydberg state.}
\subsection{DLCZ quantum memory}

\begin{figure}[h] 
\includegraphics[width=0.5\linewidth]{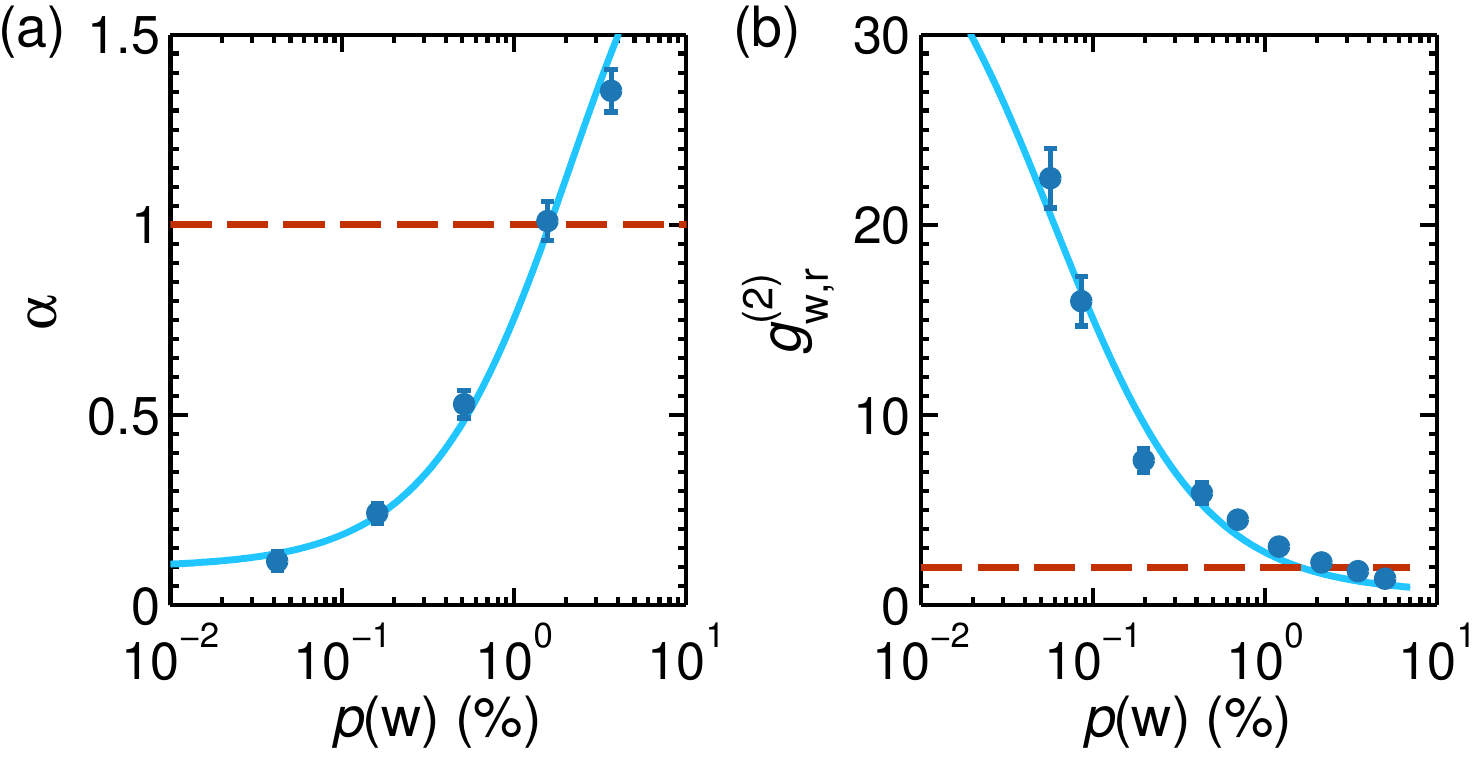}
\caption{\textbf{Anti-bunching parameter and cross-correlation $g^\mathrm{(2)}_\mathrm{w,r}$ without storage in site B}(a) Anti-bunching parameter $\alpha$ measured before site B and (b) cross-correlation function $g^\mathrm{(2)}_\mathrm{w,r}$ measured after site B without loading the atomic ensemble. Data are taken at $t_{\rm A} \sim 1 \mu$s. For low $p({\rm w})$ high quality heralded single photon in the read mode as well as non-classical correlations are created, beating the classical bounds (indicated by the dashed line). The solid lines are fits with a model described in Supplementary Note 2. The error bars are the propagated Poissonian error of the photon counting probabilities.}
\label{fig:fig2}
\end{figure}

First, at fixed $t_{\rm A} \sim 1 \, \mu$s we characterize the DLCZ memory in site A as a source of high-quality synchronizable single photons when no storage in site B is performed, as shown in Figure \ref{fig:fig2}. The read single photon quality is inferred by measuring its heralded anti-bunching parameter $\alpha=p({\rm r}_3,{\rm r}_4|{\rm w})/p({\rm r}_3|{\rm w})p({\rm r}_4|{\rm w})$ via Hanbury Brown-Twiss (HBT) measurement before the $10 \,\rm m$ SMF. We also measure the second-order, cross-correlation function $g^\mathrm{(2)}_{\mathrm{w,r}} = p_0({\rm w},{\rm r}_2)/p({\rm w})p_0({\rm r}_2)$ of the paired write and read photons without loading the atoms in site B. Here $p({\rm w})$ ($p({\rm r}_i)$) is the probability to detect a write (read) photon by SPD D1 (D$i$, with $i=2,3,4$), while $p(x,y)$ is the probability of coincident detection event $x$ and $y$ and $p(x|y)$ is the conditional probability of event $x$ conditioned on $y$. The subscript $0$ indicates 
that no atoms are loaded in site B. At low $p({\rm w})$, a successful detection of a write photon projects the read mode into a high-quality single-photon state, with measured values as low as 
$\alpha = 0.11 \pm 0.02$ at $p({\rm w}) = 0.04 \,\%$, shown in Figure \ref{fig:fig2}a. In the same condition strong non-classical correlations are found, $g^\mathrm{(2)}_\mathrm{w, r}$ being well above the classical bound of 2 for a state emitted by a DLCZ QM ({assuming thermal statistics for the write and read fields}, see Supplementary Note 1). At higher $p({\rm w})$, multiple excitations are created in the atomic ensemble and the classical bounds for $\alpha$ and for $g^\mathrm{(2)}_\mathrm{w,r}$ are recovered (see Supplementary Note 1).

\subsection{Storage in the Rydberg ensemble}

\begin{figure}[h]
 \includegraphics[width=0.5\linewidth]{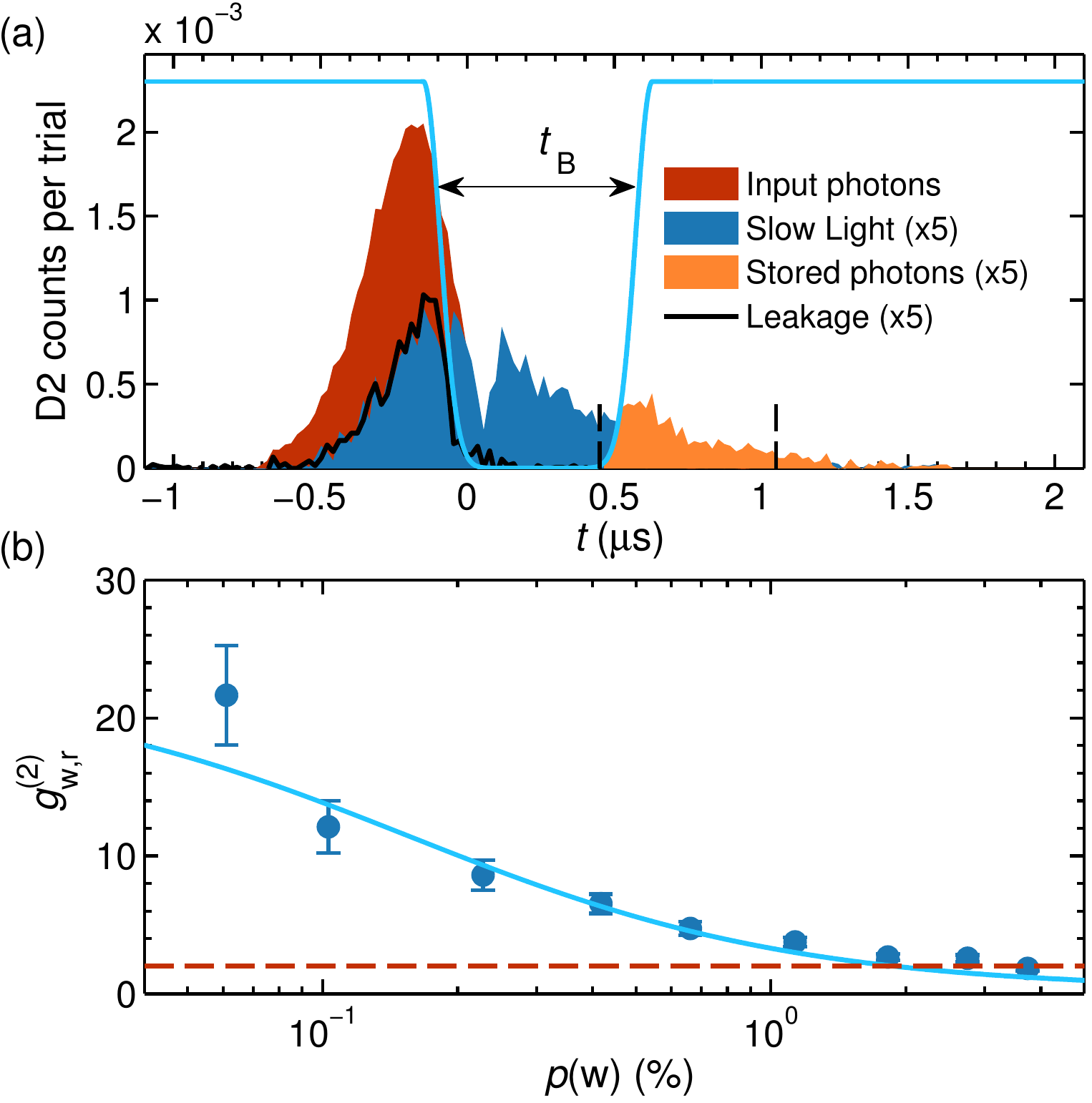}
 \caption{\textbf{Single-photon storage example and cross-correlation} (a) Example of single-photon storage for $t_{\rm A} \sim 1 \, \mu$s and $p(w) = 2.7 \,\%$. Detected counts of single-photon detector D2 per trial and per temporal bin, conditioned on a detection of a write photon, as a function of the detection time $t$ when no atoms are loaded in site B (red area), when the read photon is slowed by the presence of the coupling beam (i.e. when the coupling beam is kept on, blue area) and when the read photon is stored and retrieved for $t_{\rm B} = 500 \,$ ns (orange area). We attribute the dip at $t\sim 0 \,\mu$s observable in the slow light pulse to the fast switch-off of the trailing edge of the input read photon (see \cite{Wei2009,Zhang2011}). The solid black line represent a leakage of the slowed read photon due to low optical depth of the ensemble in site B. The solid light blue line is a pictorial representation of the coupling beam power. The vertical dashed lines shows the $600 \, $ns temporal window chosen for measuring $p({\rm w},{\rm r}_2)$. In this example the storage efficiency is $\eta_{\rm B} = 3.4 \, \%$. We refer the reader to Supplementary Note 5 and Supplementary Figure 4 for a description of the cross-correlation function across the slowed-down pulse. (b) $g^\mathrm{(2)}_\mathrm{w,r}$ as a function of $p({\rm w})$ after storage and retrieval of the read photon for $t_{\rm B} = 500 \,$ns. The error bars represent the propagated Poissonian error of the photon counting probabilities. The solid line is a fit with a model given in Supplementary Note 2 from which we extract the intrinsic retrieval efficiency of the DLCZ source $\eta_{\rm A} = 38.5 \, \%$. Dashed horizontal line shows the classical bound $g^\mathrm{(2)}_\mathrm{w,r}=2$.}  
 \label{fig:g2vsPw}
\end{figure}

We then store the emitted single photon in a collective high-lying Rydberg atomic excitation (see Figure \ref{fig:g2vsPw}a). Keeping a fixed $t_{\rm A} \sim 1 \, \mu$s, we load the atoms in site B and we store the read photon as atomic coherence between states $\ket{g_{\rm B}}$ and $\ket{R_{\rm B}}$ by switching off the coupling beam while the photon is propagating through the ensemble. After a storage time $t_{\rm B}$, we retrieve the stored excitation by switching the coupling beam back on. At $t_{\rm B}=500$ ns, we achieve a storage and retrieval efficiency of $\eta_{\rm B} = 3.4 \pm 0.4 \%$ where $\eta_{\rm B}$ is defined as $\eta_{\rm B} = p({\rm r}_2|{\rm w})/p_0({\rm r}_2|{\rm w})$. We also measure $g^\mathrm{(2)}_\mathrm{w,r} = p({\rm w},{\rm r}_2)/p({\rm w})p({\rm r}_2)$ after storage and retrieval (see Figure \ref{fig:g2vsPw}b). Our data show that $g^\mathrm{(2)}_\mathrm{w,r}\gg2$ for low $p({\rm w})$ demonstrating the persistence of non-classical correlations between the write photon and the collective Rydberg atomic excitation after storage. At $t_{\rm B}=500$ ns, we explicitly violate the Cauchy-Schwarz (CS) inequality by three to four standard deviations (see Tab. \ref{tab}), which states that a pair of classical light fields must satisfy (see \cite{Kuzmich2003}) $R = \left[ g^\mathrm{(2)}_\mathrm{w,r} \right]^2/\left[g^\mathrm{(2)}_\mathrm{w,w}\,g^\mathrm{(2)}_\mathrm{r,r}\right] \leq 1$, where $g^\mathrm{(2)}_\mathrm{w,w}$ and $g^\mathrm{(2)}_\mathrm{r,r}$ are the unheralded second order autocorrelation functions of the write and read photon, for which a similar expression as for $g^\mathrm{(2)}_\mathrm{w,r}$ holds (see Supplementary Note 1). For the same storage time we also measured the anti-bunching parameter $\alpha_{t_{\rm B}}$ of the stored and retrieved read photon by a HBT measurement after site B and we found $\alpha_{t_{\rm B}} =1.2 \pm 0.2$ at $p({\rm w}) = 3.98 \,\%$ and $\alpha_{t_{\rm B}} = 0.0 \pm 0.35$ at $p({\rm w}) = 0.59\,\%$, the latter confirming that the single-photon statistics are preserved after storage and retrieval (see Methods). 

\begin{table}[h]
\begin{center}
\begin{tabular}{||c|c|c|c||c||}
\hline 
$p({\rm w}) [\%]$ & $g^\mathrm{(2)}_\mathrm{w,r}$ & $g^\mathrm{(2)}_\mathrm{w,w}$ & $g^\mathrm{(2)}_\mathrm{r,r}$ & $R$ \\ 
\hline 
3.7 & $1.8\pm 0.2$ & $1.90\pm 0.02$& $1.5\pm 0.3$ & $1.2\pm 0.3$ \\ 
\hline 
1.13 & $3.7\pm 0.3$ & $1.97\pm 0.03$ & $1.6\pm 0.3$ & $4.4\pm 1.0$\\ 
\hline 
0.66 & $4.7\pm 0.5$ & $2.00\pm 0.06$& $1.5\pm 0.5$ & $7.7\pm 2.6$ \\ 
\hline 
\end{tabular}
\caption{\textbf{Reported value of the $R$ parameter for the Cauchy-Schwarz inequality.} Data are taken for $t_{\rm A} = 1 \, \mu \rm s$ and $t_{\rm B} = 500 \, \rm ns$. For low $p({\rm w})$, the CS inequality is explicitly violated.} \label{tab}
\end{center}
\end{table}

\begin{figure}[h]
 \includegraphics[width=0.5\linewidth]{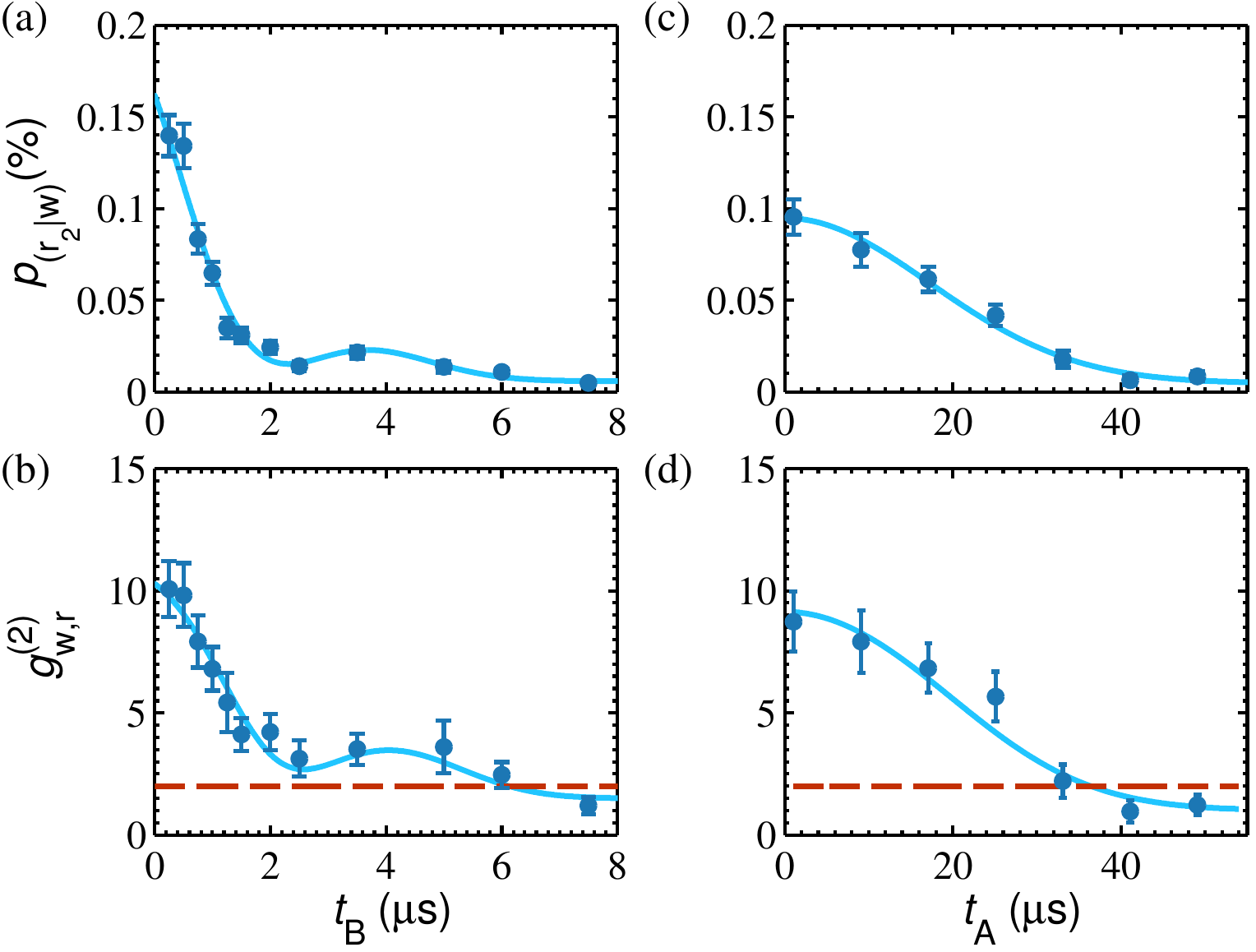}
 \caption{\textbf{Coincidence detection probability and cross-correlation functions} Coincidence detection probability $p({\rm r}_2|{\rm w})$ and $g^\mathrm{(2)}_\mathrm{w,r}$ as function of $t_{\rm B}$ for $t_{\rm A} \sim 1 \,\mu$s (panel (a) and (b)) and as a function of $t_{\rm A}$ for $t_{\rm B} = 500$ ns (panel (c) and (d)). In (a) the measured $p(r_2|w)$ at $t_{\rm B} = 500$ ns corresponds to a storage and retrieval efficiency $\eta_{\rm B} = 3.8 \pm 0.4 \%$. The solid lines are a fit with the model described in the Supplementary Notes 2 and 4, from which we extract the $1/e$ decay times of $p({\rm r}_2|{\rm w})$ being  $\tau_{\rm R} = 3.3 \pm 0.3 \, \mu \rm s$ and  $\tau_{\rm DLCZ} = 24 \pm 2 \, \mu$s for (a) and (b) respectively.  The error bars represent the propagated Poissonian error of the photon counting probabilities.} 
 \label{fig:stTime}
\end{figure} 

The memory capabilities of the Rydberg ensemble and of the DLCZ QM are studied in Figure \ref{fig:stTime} for $p({\rm w})=0.16 \pm 0.02 \,\%$. First we show $p({\rm r}_2|{\rm w})$ and $g^\mathrm{(2)}_\mathrm{w,r}$ as a function of $t_{\rm B}$ (see Figure \ref{fig:stTime}(a,b)) keeping a fixed $t_{\rm A} \sim 1\mu$s. $p({\rm r}_2|{\rm w})$, along with  $g^\mathrm{(2)}_\mathrm{w,r}$, decreases when increasing the storage time, due to atomic motion and external residual fields which dephase the collective Rydberg state of equation \eqref{CollectiveRydberg}. We also observe a oscillatory revival  which we attribute to the hyperfine splitting $\Delta F$ of the Rydberg state $\ket{R_{\rm B}}$ resulting in a beating of $p({\rm r}_2|{\rm w})$ with a period $T = 1/\Delta F$. The non-classical correlations between a photon and a stored Rydberg excitation are preserved up to around $t_{\rm B} \sim 6 \mu s$. Fitting $p({\rm r}_2|{\rm w})$ and $g^\mathrm{(2)}_\mathrm{w,r}$ with a model shown in Supplementary Notes 2 and 4, we extract the $1/e$ decay time of the storage efficiency, $\tau_{\rm R} = 3.3 \pm 0.3 \, \mu \rm s$ as well as $\Delta F = 170 \pm 16$ kHz, the latter being compatible with the theoretical value of $\Delta F_{\rm theo} = 182.3$ kHz.

We also verify that we can generate the write and the read photon with long, controllable delay in site A, maintaining the non-classical correlation between them after storage and retrieval in site B. This result is shown in Figure \ref{fig:stTime}(c,d) where we change the read-out time $t_{\rm B}$ of the stored ground-state spin-wave while keeping a fixed $t_{\rm B} = 500$ ns. Here, the ground state storage ensures a storage time longer than in the Rydberg state. In this case the $1/e$ decay time is $\tau_{\rm DLCZ} = 24 \pm 2 \, \mu$s and we observe non-classical correlations between the write and the stored and retrieved read photon in site B up to $t_{\rm A} \sim 30 \, \mu \rm s$. 

\subsection{Nonlinear response of the Rydberg ensemble}

\begin{figure}[h]
 \includegraphics[width=0.5\linewidth]{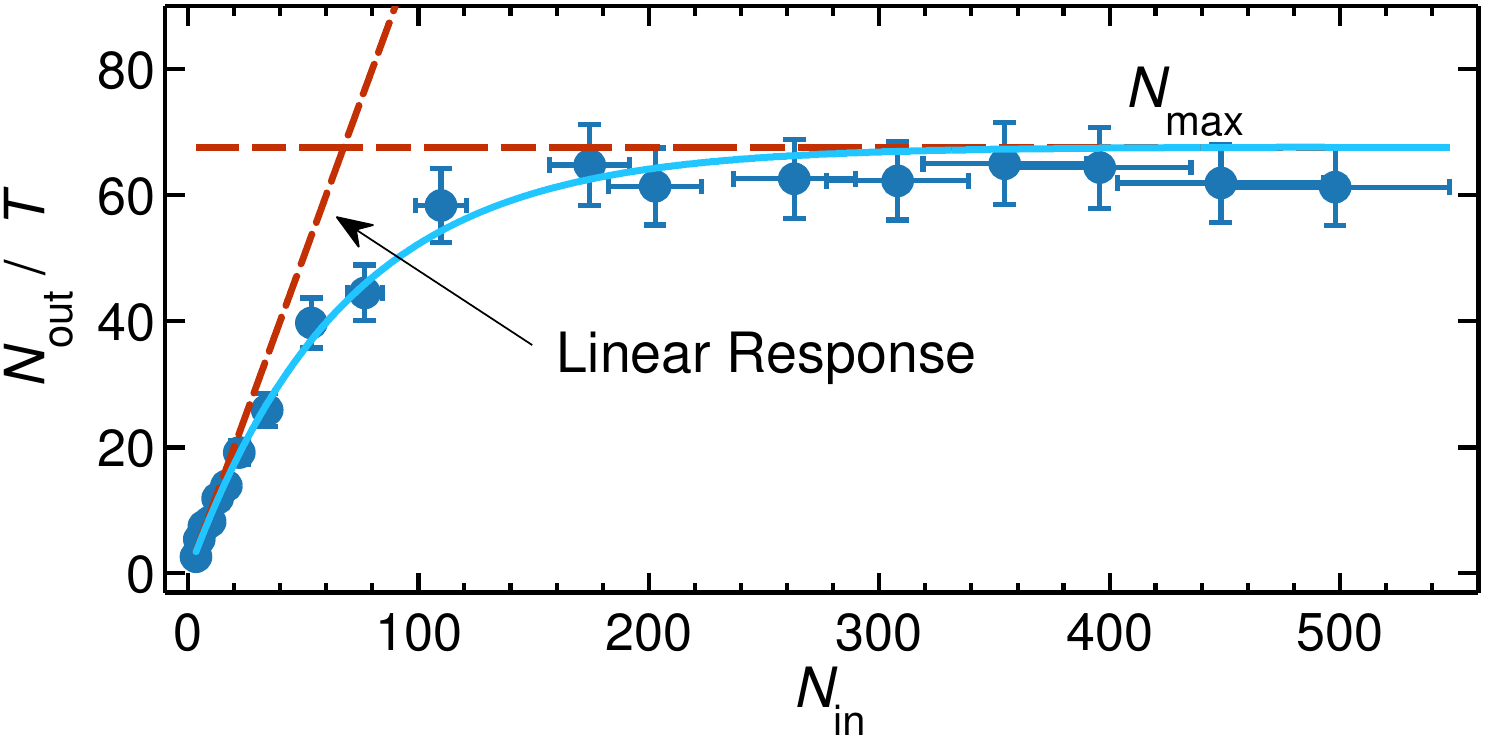}
 \caption{\textbf{Nonlinear response of the Rydberg blockaded ensemble in site B.} We store weak coherent states with a varying mean number of photon $N_{\rm in}$ into the Rydberg state $\ket{R_{\rm B}}$ for a storage time $t_{A} = 4 \,\mu$s and we measure the mean number of output photon $N_{\rm out}$. We plot $N_{\rm out}$ normalized by the storage efficiency at low number of photons $T$ as a function of $N_{\rm in}$. Due to Rydberg induced photon blockade, the medium can stand a maximum of $N_{\rm max} = 68 \, \pm \, 8$. The solid line is a fit with the model described in the Supplementary Note 4. In this example $T = 0.44 \pm 0.02 \, \% $. The error bars are the propagated Poissonian error of the photon counting probabilities.}  
 \label{fig:nonlinear}
\end{figure}

Finally we prove the highly nonlinear response of the Rydberg ensemble. This is demonstrated by storing for $4 \,\mu$s weak coherent states with varying mean number of input photon $N_{\rm in}$ and measuring the mean number of photons in the retrieved pulse after storage $N_{\rm out}$, in a way presented in \cite{Distante2016}. For a linear medium $N_{\rm out} = T N_{\rm in}$ where $T$ is the storage efficiency, while here we show (Figure \ref{fig:nonlinear}) strong nonlinear dependence. DD interactions prevent many excitations to be stored and retrieved in the medium, which can therefore sustain no more than $N_{\rm max}$ photons. As a consequence $N_{\rm out}$ becomes $N_{\rm out}= N_{\rm max} T (1-e^{-N_{\rm in}/N_{\rm max}})$ \cite{Baur2014a}. Our result shows (see Methods) $N_{\rm max} = 68 \pm 8$ although the nonlinear dependence of $N_{\rm out}$ with respect to $N_{\rm in}$ appears at a lower number of photons. It should be noted that this result is obtained with a standard magneto-optical trap with a moderate atomic density, and that this result shows a nonlinearity 6 times stronger than the one reported in \cite{Distante2016}. As demonstrated in \cite{Peyronel2012,Firstenberg2013,Maxwell2013}, increasing the density of atomic ensemble with known atomic trapping techniques will allow us to achieve nonlinearity at the single-photon level, as required for applications in quantum information science. 

To summarize, we have demonstrated for the first time storage and retrieval of a paired single photon on a highly nonlinear medium based on an atomic ensemble. The nonlinearity relies on highly excited Rydberg states where the capability of successfully storing a single photon is of particular importance for implementing high fidelity quantum gates. The source is based on an emissive QM with multimode capability \cite{Albrecht2015} which is particularly suitable for quantum networking applications. Connecting this type of source with a highly nonlinear medium represents a building block for quantum networks where the entanglement can be deterministically shared over long distance by deterministic BSMs.

\section{Methods}
\subsection{ DLCZ ensemble}
In site A the measured optical depth is $\rm OD \sim 5$ on the $\ket{g_{\rm A}}\rightarrow \ket{e_{\rm A}}$ transition. A bias magnetic field $B = 110$ mG along the read and write photon direction defines the quantization axes. Write and read pulses are opposite circularly polarized $\sigma_{-}$ and $\sigma_{+}$ respectively. The write pulses have a Gaussian temporal shape of duration FWHM $\sim 20 \, \rm ns$. The power and the temporal shape of the read pulses have been tailored to optimize the signal-to-noise ratio of the stored and retrieved read photon which results in a Gaussian shape of FWHM $\sim 350 \, \rm ns$. The angle between the write/read pulses and the write/read photon is $\theta = 3.4º$.  From $\theta$ and from the Gaussian decay time $\tau_{\rm DLCZ}$ extracted from the fit shown in Figure~\ref{fig:stTime}(c-d), we calculate an atomic temperature of $T_{\rm A} = 77 \,\mu$K (see Supplementary Note 1). An optical cavity of finesse $F = 200$ resonant with the write photons is used in front of detector D1 as a frequency filter in combination with a polarizing beam splitter, a quarter-wave plate and a half-wave plate that serve as polarization filtering. 
\subsection{ Rydberg ensemble}
In site B, the measured OD on the $\ket{g_{\rm B}}\rightarrow \ket{e_{\rm B}}$ is OD $\sim 5.5$. The Rabi frequency of the coupling beam is $\Omega_{\rm c} = 2.66 \pm 0.06$ MHz which results in a width of the EIT line of FWHM $\sim 0.7 $ MHz. The magnetic field is nulled via microwave spectroscopy. The read photon and the coupling beam are focused to waists radii $(w_{\rm r},w_{\rm c}) \sim (7,13) \, \mu$m respectively. From the Gaussian decay $\tau_{\rm R}$ we extracted an atomic temperature of $T_{\rm B} = 38 \,\pm \, 6 \mu$K (see Supplementary Note 4). The overall detection efficiency including fiber coupling losses and efficiency of the SPD D2 is $\eta_{\rm det} = 15.2 \%$
\subsection{Measurement of coincidences}
To measure coincidences, we use a temporal window of $600$ ns around the stored and retrieved read photon and a temporal window of $60$ ns around the detected write photon. The measured value of the anti-bunching parameter after $t_{\rm B} = 500$ ns, $\alpha_{t_{\rm B}} =0.00 \pm 0.35$, at $p(w) = 0.59\,\%$ corresponds to zero counts in the coincidence windows after 19 hours of data acquisition.
\subsection{Single-photon Rydberg memory linewidth}
The single-photon Rydberg memory linewidth is set by the width of the EIT line in combination with the single photon bandwidth. In Supplementary Figure 5 we show $p(r_2|w)$ after storage and retrieval of the read photon for $t_{\rm B} = 500\,\rm ns$ in the Rydberg state $\ket{R_{\rm B}}$ as a function of the coupling beam detuning $\delta_c$ with respect the transition $\ket{e_{\rm B}}\rightarrow \ket{R_{\rm B}}$. We fit the result with a Gaussian function and we extract a width of $\rm FWHM = 2.38 \pm 0.09\,\rm MHz$ which is the convolution of the EIT linewidth and the read photon spectral width. From the measured EIT linewidth (FWHM$_{\rm EIT} = 730 \unit{kHz}$) we find a read photon spectral width of $\rm FWHM_{\rm r} = \sqrt{\rm FWHM^2 - \rm FWHM^2_{\rm EIT}} = 2.26 \pm 0.09 \,\rm MHz$. This proves that the heralded read photon can be generated in a DLCZ scheme with sub-natural linewidth in a given temporal mode, as demonstrated in \cite{Farrera2016}. Still, the spectral width of the read photon is slightly larger than the Fourier transform of its duration. We attribute this discrepancy to the long term laser drift.  
\subsection{Characterization of the nonlinearity}
Data for Figure \ref{fig:nonlinear} are taken with an increased Rabi frequency of the coupling $ \Omega_{\rm c} = 4.7 \pm  0.1 \, \rm MHz$ which results in a width of the EIT window of FWHM $= 1.3 \pm 0.04 \, \rm MHz$. Still the storage efficiency at low photon number $T$ decreased with respect to data in Figure \ref{fig:stTime}. We attribute this decrease to an external unwanted stray electric field which fluctuates during time.
\subsection{Measurement of the cross-correlation function}
As shown in Supplementary Figure 1, we build a start-stop histogram where the start is a write photon detection and the stop is a read photon detection and we measure the number coincidence detection events in the SPDs D1 and D2, $C_{\rm D1,D2}$. We then compare $C_{\rm D1,D2}$ with the coincidences due to accidentals uncorrelated detection, $C^{\rm(acc)}_{\rm D1,D2}$. To measure $C_{\rm D1,D2}$, we count the coincidences in a 60 ns long temporal window in D1 and in a 600 ns long temporal window in D2. The two detection windows are temporally separated by a time $t_{\rm A}+t_{\rm B}$, to take into account the storage time in the two ensembles. $C^{\rm (acc)}_{\rm D1,D2}$ is measured by counting the coincidences between a fist write photon and a read photon detection coming from a successive uncorrelated trial. We then measure the cross-correlation via:
\begin{equation}
g_{\rm w,r}^{(2)} = \frac{C_{\rm D1,D2}}{\langle C^{\rm(acc)}_{\rm D1,D2}\rangle},
\end{equation}      
where $\langle C^{\rm(acc)}_{\rm D1,D2}\rangle$ is the average number of coincidences in the extra trials, i.e. second to seventh peak in Supplementary Figure 1.
\subsection{Data availability}
The data appearing in Figures 2, 3b and 4 are available in Zenodo with the identifier doi:10.5281/zenodo.165760\cite{Zenodo_Distante_Farrera}. Other data may be available upon reasonable request.

\section{Acknowledgements}
We acknowledge financial support by  the ERC starting grant QuLIMA, by the Spanish Ministry of Economy and Competitiveness (MINECO) through grant FIS2015-69535-R (MINECO/FEDER) and  Severo Ochoa  SEV-2015-0522, by AGAUR via 2014 SGR 1554 and by Fundaci\'{o} Privada Cellex. 
DPB has received funding from the European Union’s Horizon 2020 research and innovation programme under the Marie Skłodowska-Curie grant agreement No 658258. P.F. acknowledges the International PhD-fellowship program ”la Caixa”-Severo Ochoa @ ICFO. G.H. acknowledges support by the ICFOnest+ international postdoctoral fellowship program
\section{Author Contribution}
All authors contributed to all aspects of this work.
\section{Competing financial interests}
The authors declare no competing financial interests.




\newpage

\section*{Supplementary Notes}
\subsection*{Supplementary Note 1: Theoretical background of a DLCZ quantum memory}
In this section we review the basics of a DLCZ quantum memory. In site A, $N_{\rm A}$ atoms are prepared in the ground state, and their collective state is written as $\ket{G}=\ket{g_{{\rm A}_1} ... g_{{\rm A}_i} ... g_{{\rm A}_{N_{\rm A}}}}$. A write pulse generates with a probability $p$ at least a write photon in a given spatial mode by Raman scattering, transferring the atoms in the ground state $\ket{s_{\rm A}}$. The atom-light system can be described by the two-mode squeezed state: 
\begin{equation}\label{two-modesqueezedstate}
\ket{\psi_A} = \sqrt{1-p}\sum_{n=1}^{\infty} p^{n/2}\frac{\left(a_{\rm w}^\dag S^\dag\right)^n}{n!} \ket{0_{\rm w}} \ket{G}. 
\end{equation}
Here the subscript $\rm w$ indicates the write photon mode, $a_{\rm w}^\dag$ and $S^\dag$ are the creation operators of the write photon and of the collective state of the atoms, explicitly:
\begin{equation}
S^\dag=\frac{1}{\sqrt{N_A}}\sum_{j=1}^{N_A} e^{-i(\kW-\kw)\cdot\bold{r}_j }\ket{s_{{\rm A}_j}}\bra{g_{{\rm A}_j}}, 
\end{equation}
where $\kW$ and $\kw$ are the wavevectors of the write pulse and the write photon, and $\bold{r}_j$ is the position of the $j^{\rm th}-$atom. From equation \eqref{two-modesqueezedstate}, one can see that for low $p$ the atoms share a single, collective spin excitation (also called \emph{spin-wave})

After a storage time $t_{\rm A}$ the atomic excitation is mapped deterministically into a read photon via Raman scattering by mean of an intense read pulse. Defining the read pulse wavevector as $\kR$, the read photon spatial-mode will be given by the phase-matching condition $\kr=\kW+\kR-\kw$, while the temporal mode will depend on the temporal shape of the read pulse \cite{Farrera2016}. The pair state of the write and read photon then reads:   

\begin{equation}\label{pair_state}
\ket{\varphi} = \sqrt{1-p}(\ket{0_{\rm w} 0_{\rm r}} + \sqrt{p}\ket{1_{\rm w} 1_{\rm r}} + p \ket{2_{\rm w} 2_{\rm r}} + O(p^{3/2})),
\end{equation}

where now the subscript $\rm r$ stands for the read photon mode. From equation \eqref{pair_state}, it is evident that for low $p$ a successful detection of a write photon project the read photon into a single photon state.

The write and read photon statistics can be measured by their second-order auto-correlation function $g_{\rm w,w}^{(2)}$ and $g_{\rm r,r}^{(2)}$ via HBT measurement. Taking as example the write photon mode (similar expression holds for the read photon). we denote ${\rm w}_1$ and ${\rm w}_{2}$ the write photon mode after passing through a balance beam splitter, $p({\rm w}_i)$ the probability to detect a single photon in the ${\rm w}_{i}$ photon mode, and $p({\rm w}_1,{\rm w}_2)$ the probability of a coincidence detection of photons between the modes ${\rm w}_{1}$ and ${\rm w}_2$, the auto-correlation function reads:
\begin{equation}\label{g2ww}
g_{\rm w,w}^{(2)} = \frac{p({\rm w}_1,{\rm w}_2)}{p({\rm w}_1)p({\rm w}_2)} = \frac{\langle a_{\rm w_1}^\dag a_{\rm w_2}^\dag a_{\rm w_2} a_{\rm w_1} \rangle}{\langle a_{\rm w_1}^\dag a_{\rm w_1}\rangle \langle a_{\rm w_2}^\dag a_{\rm w_2} \rangle}.
\end{equation}
   
From Eq. (\ref{g2ww}) one finds $g_{\rm w,w}^{(2)} = g_{r,r}^{(2)} = 2$ for the two-mode squeezed state given by equation \eqref{pair_state}. This means that taken individually the write and the read photon are in a pure thermal state. Stray light and coherent leakage noise diminishes the auto-correlation function thus one finds $1 \leq g_{\rm w,w}^{(2)}\leq 2$ and $1 \leq g_{\rm r,r}^{(2)}\leq 2$, as also shown by our measurement (see Table I in the main text).

On the contrary, after a successful detection of a write photon, for low $p$ the read photon mode is projected onto a single photon states, showing anti-bunching in a HBT measurement. That means that its \emph{heralded} auto-correlation function, also called the anti-bunching parameter $\alpha=g_{(r,r|w)}^{(2)}$, must be $\alpha \ll 1$. Explicitly:
\begin{equation}
\alpha = \frac{p({\rm r}_1,{\rm r}_2|{\rm w})}{p({\rm r}_1|{\rm w})p({\rm r}_2|{\rm w})} = \frac{\langle a_{\rm w}^\dag a_{\rm r_1}^\dag a_{\rm r_2}^\dag a_{\rm r_2} a_{\rm r_1} a_{\rm w} \rangle \langle a_{\rm w}^\dag a_{\rm w}\rangle}{\langle a_{\rm w}^\dag a_{\rm r_1}^\dag a_{\rm r_1} a_{\rm w} \rangle \langle a_{\rm w}^\dag a_{\rm r_2}^\dag a_{\rm r_2} a_{\rm w} \rangle}, 
\end{equation}
where it can be shown that 
\begin{equation}\label{alpha:eq}
\alpha = 2p\frac{(2+p)}{(1+p)^2},
\end{equation}
and $\alpha \underset{p \rightarrow 0 }{\longrightarrow} 4p$.

The correlations between the write and the read photon can be measured by the second-order cross-correlation function $g_{\rm w,r}^{(2)}$, which as a function of $p$ is:

\begin{equation}
g_{\rm w,r}^{(2)} = \frac{\langle a_{\rm w}^\dag a_{\rm r}^\dag a_{\rm r} a_{\rm w} \rangle}{\langle a_{\rm w}^\dag a_{\rm w}\rangle \langle a_{\rm r}^\dag a_{\rm r}\rangle} = 1 + \frac{1}{p}
\end{equation}

from which it is clear that for high-quality single photon (i.e. $\alpha \sim 0$) one gets $g_{\rm w,r}^{(2)} \gg 1$.

Non-classical correlations between the write and the read photons arise when $g_{\rm w,r}^{(2)}>2$. This can be inferred from the Cauchy-Schwartz inequality (CS) which states that two classical light fields fulfil $R = \left[ g_{\rm w,r}^{(2)} \right]^2/g_{\rm w,w}^{(2)}g_{\rm r,r}^{(2)} \leq 1$. Since in our situation  $1 \leq g_{\rm w,w}^{(2)}\leq 2$ and $1 \leq g_{\rm r,r}^{(2)}\leq 2$ for the write and the read photon, it is easy to see that $g_{\rm w,r}^{(2)}>2$ violates the CS inequality. 

\subsection*{Supplementary Note 2 Description of the photon emission from the DLCZ photon source including noise}
Here we show the theoretical model that we use to fit data in Fig. 2(a,b), 3b, 4(a,b,c,d) in the main text. Following a similar procedure as in \cite{Farrera2016a}, we start from the ideal description of a DLCZ QM and we include the noise contributions given by spontaneous emission, stray light fields and dark-counts of the SPDs.    

As explained in detail above, the write process creates pairs consisting of a write photon and a spin-wave in a probabilistic way. These spin-waves can be later converted with a high probability into read photons emitted in a particular transition and direction, due to collective interference. The quantities that we measure in this work depend on how directional is the emission of these read photons. The ratio between directional and random emission depends, among other factors, on the preservation of the spin-wave coherence or the optical depth of the atomic ensemble. 

The ratio of directional emission can be characterized by the intrinsic retrieval efficiency $\eta_{\rm A}$. The random emission is proportional the total number of atoms in the $\ket{s_{\rm A}}$ ground state, the read photon spatial mode solid angle and to global branching ratio corresponding to the detectable transitions, $p_{\rm SE}$. Following a similar procedure as in \cite{Farrera2016a}, we can write all the photon detection probabilities as:

\begin{align}
p({\rm w})&= p\eta_{\rm w}+p_{n{\rm w}}, \label{eq:1}\\
p({\rm r})&= p({\rm r})^{dir}+p({\rm r})^{\rm rand}=\nonumber \\ 
    &= p\eta_{\rm A}\eta_{{\rm r}}+p(1-\eta_{\rm A})p_{\rm SE}\eta_{\rm r}+p_{n{\rm r}}, \label{eq:2}\\
p({\rm w},{\rm r})&=p({\rm w},{\rm r})^{dir}+p({\rm w})p({\rm r})^{\rm rand}= \nonumber \\   
      &= p({\rm w})\eta_{\rm A}\eta_{{\rm r}}+p({\rm w})p(1-\eta_{\rm A})p_{\rm SE}\eta_{\rm r}+p({\rm w})p_{n{\rm r}}, \label{eq:3} 
\end{align} 

where $p$ is the probability to create a spin-wave together with a write photon in the coupled spatial mode, $\eta_{{\rm w}}$ is the write photon total detection efficiency, $\eta_{{\rm r}}$ is the total transmission of the read photon after leaving the atomic ensemble (including the Rydberg storage efficiency $\eta_{\rm B}$) and $p_{n{\rm w}}$ ($p_{n{\rm r}}$) is the probability to detect a background count (including stray light and SPD dark-counts) in the write (read) photonic mode.
To fit data of Fig. 2a of the main text we used equation \eqref{alpha:eq} substituting $p$ with $c_1 p({\rm w})+c_2$. The terms $c_1$ and $c_2$ are used to include the noise on the detectors D1, D3 and D4. 

From Eqs. (\ref{eq:1},~\ref{eq:2},~\ref{eq:3}) we compute the function:
\begin{equation}
g_{\rm w,r}^{(2)} = \frac{p({\rm w},{\rm r})}{p({\rm w})p({\rm r})},
\end{equation}

which is used to fit the data of Fig. 2b and 3b in the main text. The free parameters are $\eta_{\rm A}$, $p_{\rm SE}$, and $p_{nr}$. In our experiment, the SPD of the write photon is gated for a very short time $t^{gate}_{\rm w} \sim 60 \, \rm ns$, therefore we measured $p_{nw} \ll p(w)$. 

To fit data in Fig. 4(a,b) of the main text, a storage time dependence of the Rydberg storage efficiency has to be included. To do so, we considered a time-dependent $\eta_{r}$, where the time dependence follows equation \eqref{storage_efficiency} (see next section). Similarly, for the data in Figures 4(c,d) the delay between the write and read pulses is changed and the decoherence of the spin-wave has to be considered. Decoherence decreases the directionality of the read photon emission. In our experiment this is mainly due to the motion of the atoms and it gives a Gaussian decay of the intrinsic retrieval efficiency $\eta_{\rm A}(t)=\eta_{\rm A}e^{-t^2/\tau_{\rm DLCZ}^2}$ \cite{Albrecht2015}, being $\tau_{\rm DLCZ} = \sqrt{m/(k_BT\Delta k^2)}$, where $m$ is the atomic mass, $k_{\rm B}$ is Boltzmann's constant, $T$ is the atomic temperature and $\Delta k=|\kW-\kw|$ is the difference between the write pulse and write photon wavevectors. 

\subsection*{Supplementary Note 3 Theoretical background of electromagnetically induced transparency}
Here we review briefly the basics of the electromagnetically induced transparency (EIT) in site B. The atomic cloud in site B can be described as a system of three level atoms, the levels being: the ground state $\ket{g_{\rm R}} = \ket{5{\rm} S_{1/2}, F=2}$, the excited state $\ket{e_{\rm B}} = \ket{5{\rm} P_{3/2}, F=2}$ and the Rydberg state $\ket{R_{\rm B}} = \ket{60{\rm} S_{1/2}}$. We probe the atomic cloud by measuring the transmission through the cloud of a weak coherent field ($\mathcal{E}$) detuned by $\delta$ with respect to the $\ket{g_{\rm B}}\rightarrow\ket{e_{\rm B}}$ transition. When the detuning approaches the resonant condition, i.e. $\delta = 0$, the atoms absorb the probe field, the transmission drops to its minimum value $T(\delta = 0) = e^{\rm -OD}$, where $\rm OD$ is the Optical Depth of the cloud for this transition. By fitting the data of the transmission as a function of $\delta$ we extract the $\rm OD$ of the cloud, being $\rm OD = 5.41 \pm 0.14$.

A strong coupling beam with Rabi frequency $\Omega_{\rm c}$ resonant with the $\ket{e_{\rm B}}\rightarrow\ket{R_{\rm B}}$ transition creates the conditions for EIT opening up a window of transparency in the transmission spectrum of the probe field around $\delta = 0$. We extract $\Omega_{\rm c}$ as well as the dephasing rate $\gamma_{\rm gR}$ of the $\ket{g_{\rm B}}\rightarrow\ket{R_{\rm B}}$ transition by fitting the transmission as a function of $\delta$ with (see  \cite{Xiao1995}): 
\begin{align}
T = \,\,&\mathrm{exp}\lbrace-k_{\rm p} \ell \,\, Im \left[\bigchi(\delta)\right]\rbrace,\\
\bigchi(\delta) = \,\, &\mathrm{OD} \frac{\Gamma}{2 k_{\rm p} \ell} \left(\frac{\delta+ i\gamma_{\rm gR}}{(\Gamma/2 - i\delta)(\gamma_{\rm gR}-i\delta) + (\Omega_{\rm c}/2)^2} \right),
\end{align}
where $\Gamma$ is the linewidth of the $\ket{g_{\rm B}} \rightarrow \ket{e_{\rm B}}$ transition, $k_{p}$ is the probe beam wavenumber and $\ell$ is the atomic cloud length. An example of the transmission with and without the coupling beam is shown in \figurename~\ref{EIT:fig}. The dephasing $\gamma_{\rm gR}$ include the lasers linewidth, the lifetime of the $\ket{R_{\rm B}}$ state as well as the broadening induced by spurious external fields. $\gamma_{\rm gR}$ in combination with $\Omega_{\rm c}$ sets the height and the width of the EIT transparency window. The height of the peak is further limited by imperfect mode-matching between the probe beam and the coupling beam.

\subsection*{Supplementary Note 4 Light storage with EIT}
When $\delta = 0$, the coupling beam converts the probe field into a slowly propagating \emph{dark-state polariton} (DSP), described by the field operator $\phi(z,t)$. A DSP is a coherent superposition of the probe electric field $\mathcal{E}(z,t)$ and the atomic coherence between the ground and  - in our case - the Rydberg state, $\sigma_{\rm g,R}(z,t)$. For a medium with density $\rho$ the field of a DSP writes \cite{Fleischhauer2002}:
\begin{equation}\label{polariton}
\boldsymbol{\phi}(\bold{z},t) = \boldsymbol{\mathcal{E}}(\bold{z},t)\,\mathrm{cos}(\theta) - \sqrt{\rho}\,\boldsymbol{\sigma}_{\rm g,R}(\bold{z},t)e^{-\bold{\Delta k}\cdot\bold{z}} \,\mathrm{sin}(\theta) 
\end{equation}
where the mixing angle $\theta$ is related to the group index $n_{\rm gr}$ of the medium following $\mathrm{tan}^2(\theta) = n_{\rm gr}$ and $\bold{\Delta k} = \kc+\kp$, where $\kc$ ($\kp$) is the wavevector of the coupling (probe) field.  

The group index depends on $\Omega_{\rm c}$ and on the $\rm OD$ of the $\ket{g_{\rm B}}\rightarrow\ket{e_{\rm B}}$ transition. In particular $n_{\rm gr} \sim \rm OD/\Omega^2_{\rm c}$, as consequence a DSPs propagates at a reduced group velocity $v_{\rm gr} = c/n_{\rm gr}$. In the limit $\Omega_{\rm c}\rightarrow 0$, the group index $n_{\rm gr}$ goes to infinity and the group velocity of the polariton $v_{\rm gr}$ is reduced to zero. In this condition $\theta\rightarrow \pi/2$, therefore $\mathrm{cos}(\theta)\rightarrow 0$ and $\mathrm{sin}(\theta)\rightarrow 1$. This means that by switching off the coupling field it is possible to convert all the probe light into stationary atomic coherence, effectively storing the probe pulse as atomic coherence. In this situation, the state of the atomic ensemble can be written as a collective Rydberg atomic excitation
\begin{equation}\label{collective_Rydberg}
\ket{\psi_{\rm B}} = \frac{1}{\sqrt{N_{\rm B}}} \sum_{i=1}^{N_{\rm B}}e^{-i(\kc+\kp)\bold{r}_i}
\ket{g_{{\rm B}_1} ... R_{{\rm B}_i} ... g_{{\rm B}_{N_{\rm B}}}}.
\end{equation}

By switching the coupling beam back on, the light component of the polariton is restored and the pulse is retrieved out of the medium. The storage process is limited by low $\rm OD$, by finite transparency at the centre of the EIT feature and by imperfect frequency bandwidth matching between the probe pulse and the narrow band EIT window. In particular, at low OD the probe pulse is not fully compressed inside the atomic ensemble, resulting in a light leakage eventually reducing the storage efficiency. In our case $\rm OD = 5.4$ which gives a leaked part equal to $42 \, \%$ of the transmitted slowed pulse (see \figurename~\ref{storage:fig}). 

The storage efficiency decreases at longer storage time due to the dephasing of the collective Rydberg atomic state described by equation  \eqref{collective_Rydberg}. The main sources of dephasing are atomic motion given by finite temperature of the cloud, as well as external stray fields. In our experiment, we observe a Gaussian decay $\eta_{\rm B}(t_{\rm T}) =  \eta_0 e^{-t_{\rm T}^2/\tau_{\rm R}^2}$, where now $t_{\rm T} = t_{\rm B}+t_{\rm OFF}$ is the total time that an input light pulse takes to cross the atomic sample. $t_{\rm OFF}$ is a time-offset that takes into account the delay time $\delta t = v_{\rm gr}/\ell$ that results from the reduced group velocity of the DSP as well as the temporal profile of the coupling beam. In the present letter, we measure $t_{\rm T}$ as the difference between the centre-of-mass of the stored and retrieved pulse and the input pulse see \figurename~\ref{storage:fig}. By defining $f_{\rm in}(t)$ and $f_{\rm out}(t)$ the temporal shape of the input pulse and of the stored and retrieved pulse respectively, $t_{\rm T}$ reads: 
\begin{equation}\label{center of mass}
t_{\rm T} = \frac{\int f_{\rm out}(t) \, t \, dt}{\int f_{\rm out}(t)\,dt} - \frac{\int f_{\rm in}(t) \, t \, dt}{\int f_{\rm in}(t)\,dt}
\end{equation}
To fully understand data in Figures 4(a,b) of the main text and in \figurename~\ref{storage:fig}, one must include the hyper-fine structure of the Rydberg state, $\ket{R_{\rm B}} = \ket{60{\rm} S_{1/2}}$, which is composed of the two hyper-fine states $\ket{R_{{\rm B},\,F =1}} = \ket{60 S_{1/2},F=1}$ and $\ket{R_{{\rm B},\,F =2}} = \ket{60{\rm} S_{1/2},F=2}$ separated by $\Delta F_{\rm theo} = 182.1 \, \rm kHz$. Due to our laser linewidth, we cannot resolve these two states in our EIT spectroscopy, as a consequence the probe is stored in a coherent superposition of the $\ket{R_{{\rm B},\,F =1}}$ and the $\ket{R_{{\rm B},\,F =2}}$. The energy difference between these states results in a different phase evolution and the initial phase difference is recovered every $T=1/\Delta F_{\rm theo}$. This produces the oscillations of the storage efficiency as a function of time which then reads:

\begin{equation}\label{storage_efficiency}
\eta_{\rm B}(t_{\rm B}) = \eta_0e^{-t_{\rm T}^2/\tau^2_{\rm R}}\abs{p_{F=1}+(1-p_{F=1})e^{-2\pi \im \Delta F t_{\rm T}}}^2,
\end{equation}

where $p_{F=1}$ is the probability to excite the $\ket{R_{{\rm B},\,F=1}}$ state. The decay time $\tau_{\rm R}$ can be used to extract an upper bound of the atomic temperature. Considering the atomic motion as the only source of dephasing then the coherence time is $\tau_{\rm R} = \sqrt{m/(k_{\rm B}T\Delta k^2)}$ where $T$ is the temperature of the atoms.

\subsection*{Supplementary Note 5 Time-dependent slow-light cross-correlation function}
In this section we will show the cross-correlation function between the write and the read photon when the read photon undergoes Rydberg EIT in site B.

To understand our data (shown in \figurename~\ref{g2eit:fig}) it is crucial to analyse the frequency component of the noise carried by the input read photon. During the reading process in site A, a spontaneous emitted photon can be generated either in the $\ket{5{\rm} S_{1/2}, F=1} \leftrightarrow \ket{5{\rm} P_{3/2}, F=2}$ or in the $\ket{5{\rm} S_{1/2}, F=2} \leftrightarrow \ket{5{\rm} P_{3/2}, F=2}$ transitions. This is mainly due to imperfect optical pumping during the initialization of the DLCZ memory which let some unwanted population in the  $\ket{5{\rm} S_{1/2}, F=1}$ state. Such spontaneous emitted photons are not correlated with the write photon, therefore act as a source of noise, eventually limiting the $g^{(2)}_{\rm w,r}$. 

When the read photon is stored as a collective Rydberg state, the Rydberg EIT memory acts as a beneficial frequency noise filter. In this case, only a light pulse resonant with the $\ket{5{\rm} S_{1/2}, F=2} \rightarrow \ket{5{\rm} P_{3/2}, F=2}$ transition can be stored and retrieved. Since the ensemble in site B is prepared in the $\ket{5{\rm} S_{1/2}, F=2}$ state, an out-of-resonant spontaneous emitted photon would not interact with the atoms, leaving the ensemble in the same temporal mode as the input read photon. In general, any out of resonance noise of the input read photon can not be seen in the time window of the retrieved pulse.

This is not the case in a slow-light experiment. In \figurename~\ref{g2eit:fig}a we show the input pulse, the noise on the $\ket{5{\rm} S_{1/2}, F=1} \rightarrow \ket{5{\rm} P_{3/2}, F=2}$ (measured by loading the ensemble in site B and not performing EIT) together with the slow read photon with and without noise subtraction. As one can see, the slowed pulse and the input read pulse are not temporally separated. As a result, measuring the $g^{(2)}_{\rm w,r}$ of the write and read slowed pulse using as a coincidence temporal window the whole duration of the slowed pulse would lead to $g^{(2)}_{\rm w,r}<2$, since it would include all the out-of-resonance frequency noise components. To show this effect, we measured the $g^{(2)}_{\rm w,r}$ as a function of the time $t_{\rm w}$ at which we centered a $123 \unit{ns}$ coincidence detection window. The result is plotted in \figurename~\ref{g2eit:fig}b, where we show the cross-correlation function for three different values of $p(w)$. As one can see, in the region of time where the slowed pulse and the noise are present we found $g^{(2)}_{\rm w,r} \sim 2$. This is a combination of the uncorrelated noise, for which we expect $g^{(2)}_{\rm w,r} \sim 1$ and the non-classical correlated slowed read pulse, for which $g^{(2)}_{\rm w,r}>2$. At longer time, while the noise is not present anymore, not being slowed down, we still have the signal of the slowed read pulse, therefore the non-classical correlations are recovered. 
\newpage
\section*{Supplementary Figures}
\begin{figure}[H]
	\centering
	\includegraphics[width=9cm]{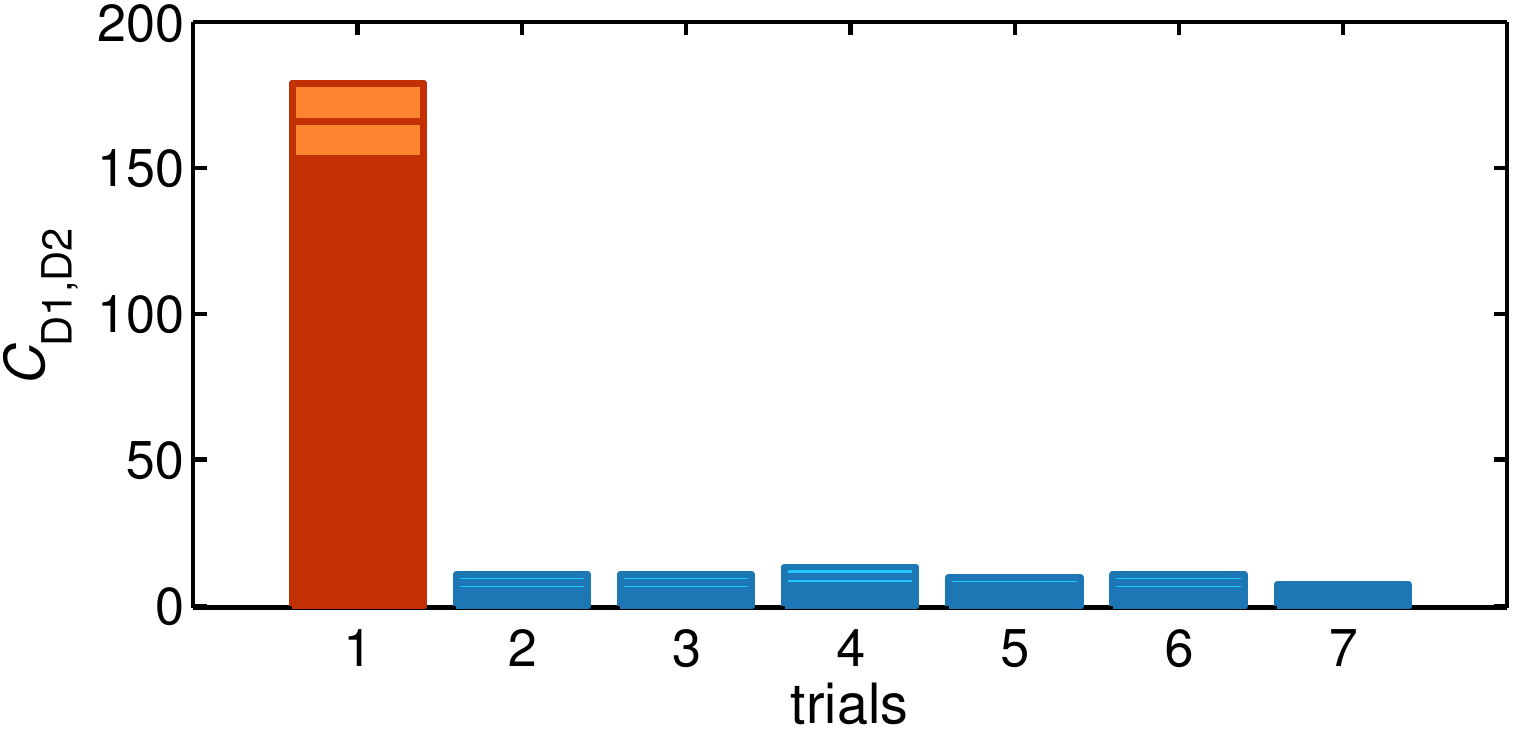}
	\caption{Number of coincidences detection events in the SPDs D1 and D2 as a function of the number of trials. The first peak (red) represents the coincidences detection between a write and a read photon proceeding from the same readout trial $C^{(1)}_{\rm D1,D2}$, while the blue peaks are the coincidences between a write photon and a read photon proceeding from a successive uncorrelated trial. The light area in each peak represents the Poissonian measurement uncertainty.}
	\label{g2histo}
\end{figure}
\begin{figure}[H]
	\centering
	\includegraphics[width=9cm]{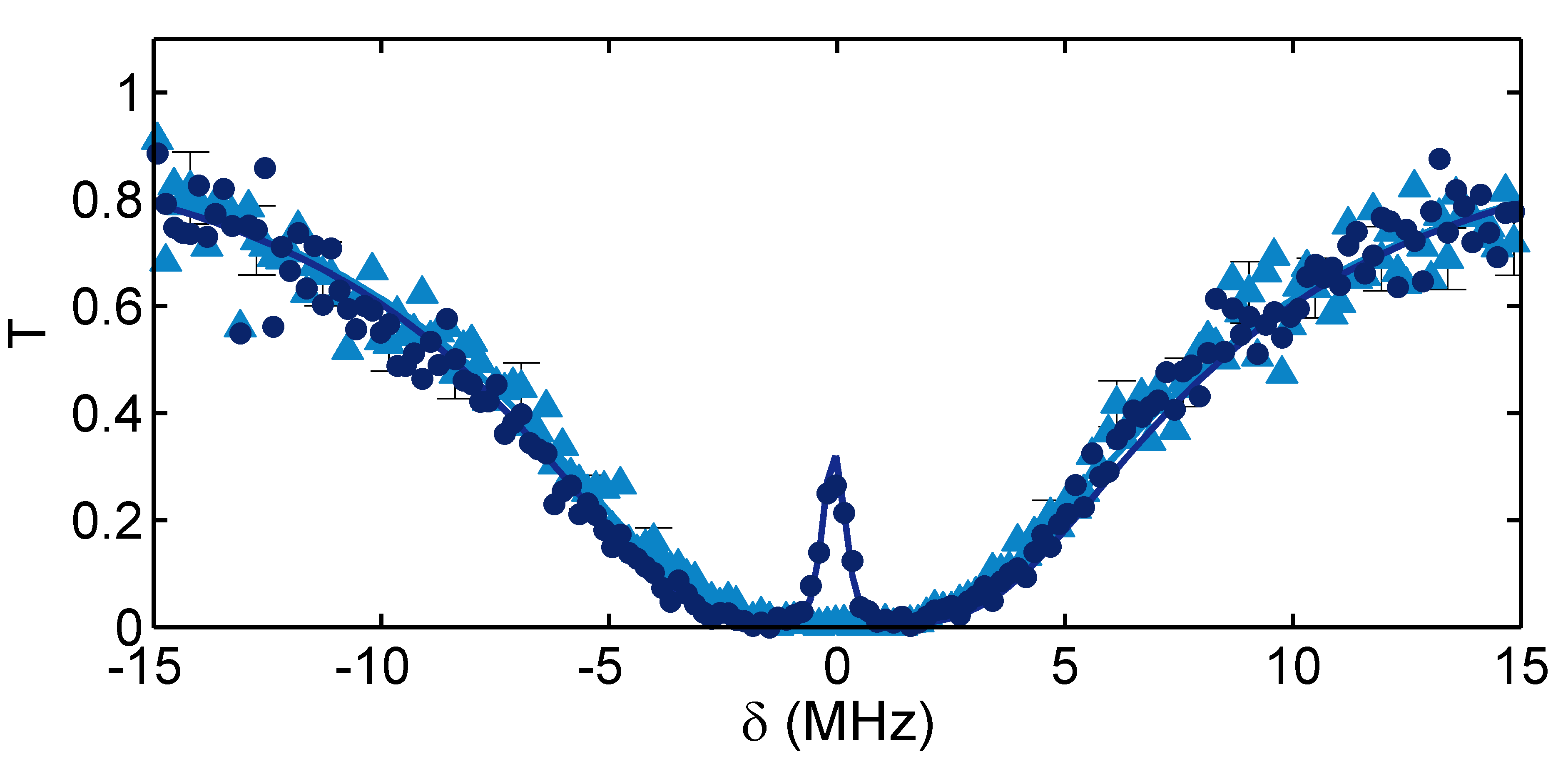}
	\caption{Transmission of the probe weak coherent state as a function of its detuning $\delta$ when the coupling beam is off (light blue triangles) and when it is on (blue circles). The solids lines of the respective colours represent a fit to the data with the model described in \cite{Xiao1995} from we extract OD = $5.4 \pm 0.1$, $\Omega_{\rm c} =  2.66 \pm 0.06 \, \rm MHz$, full width half maximum of the EIT peak FWHM = $0.73\pm 0.03 \, \rm MHz$ peak transparency $T_0=32.1 \%$ and dephasing rate $\gamma_{\rm gR} = 0.29 \pm 0.03 \, \rm MHz$. }
	\label{EIT:fig}
\end{figure}
\begin{figure}[H]
	\centering
	\includegraphics[width=9cm]{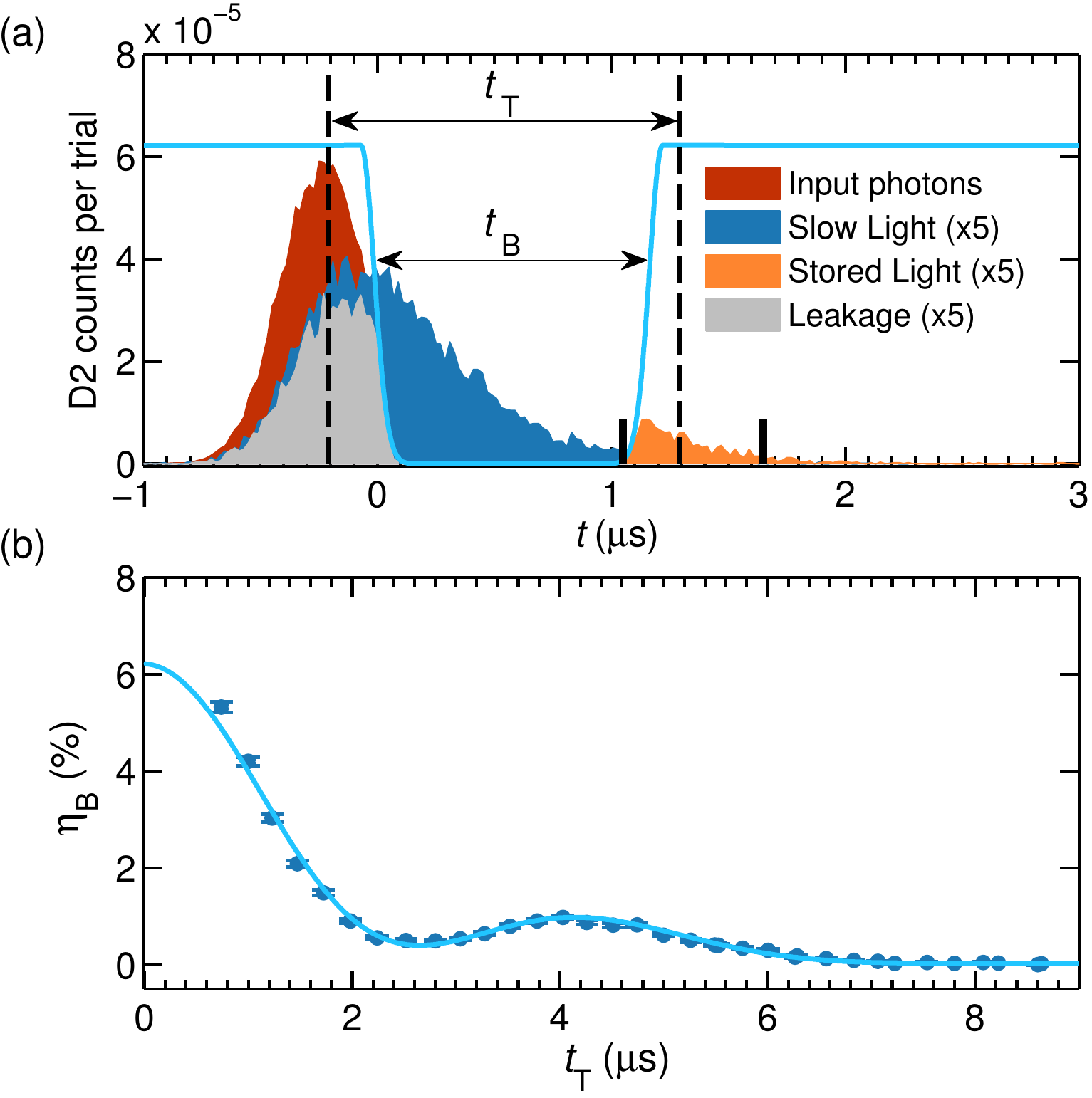}
	\caption{(a)Example of storage and slow light of a weak coherent state. Here we show: the input probe pulse measured when the atomic ensemble is not loaded (red), slowed pulse transmitted through the cloud when the coupling beam remains on (blue), the stored and retrieved pulse (orange) and the leakage due to low OD (grey). The ratio between the area of the slowed pulse and the input pulse is $\eta_{\rm slow} = A_{\rm slow}/A_{\rm in} = 23 \,\%$.  The vertical dashed line indicate the centre of mass of the input and of the retrieved pulse respectively which we use to measure $t_{\rm T}$ (see text). We calculate the storage efficiency $\eta_{\rm B}$ considering a time window of 600 ns, indicated by the black vertical lines. In this example, $\eta_{\rm B} =2.2\, \%$ for $t_{\rm B} = 1 \, \mu$s corresponding to $t_{\rm T} \sim 1.47 \, \mu$s. The solid blue line represents the intensity of the coupling beam. (b) Storage efficiency as a function of $t_{\rm T}$ together with a fit with the equation \eqref{storage_efficiency}. From the fit shown we extract a coherence time $\tau_{\rm R} = 3. 34 \pm 0.02 \, \rm ns$ and a frequency separation of the hyperfine states $\Delta F = 194 \pm 4 \, \rm kHz$.} 
	\label{storage:fig}
\end{figure}
\begin{figure}[H]
	\centering
	\includegraphics[width=9cm]{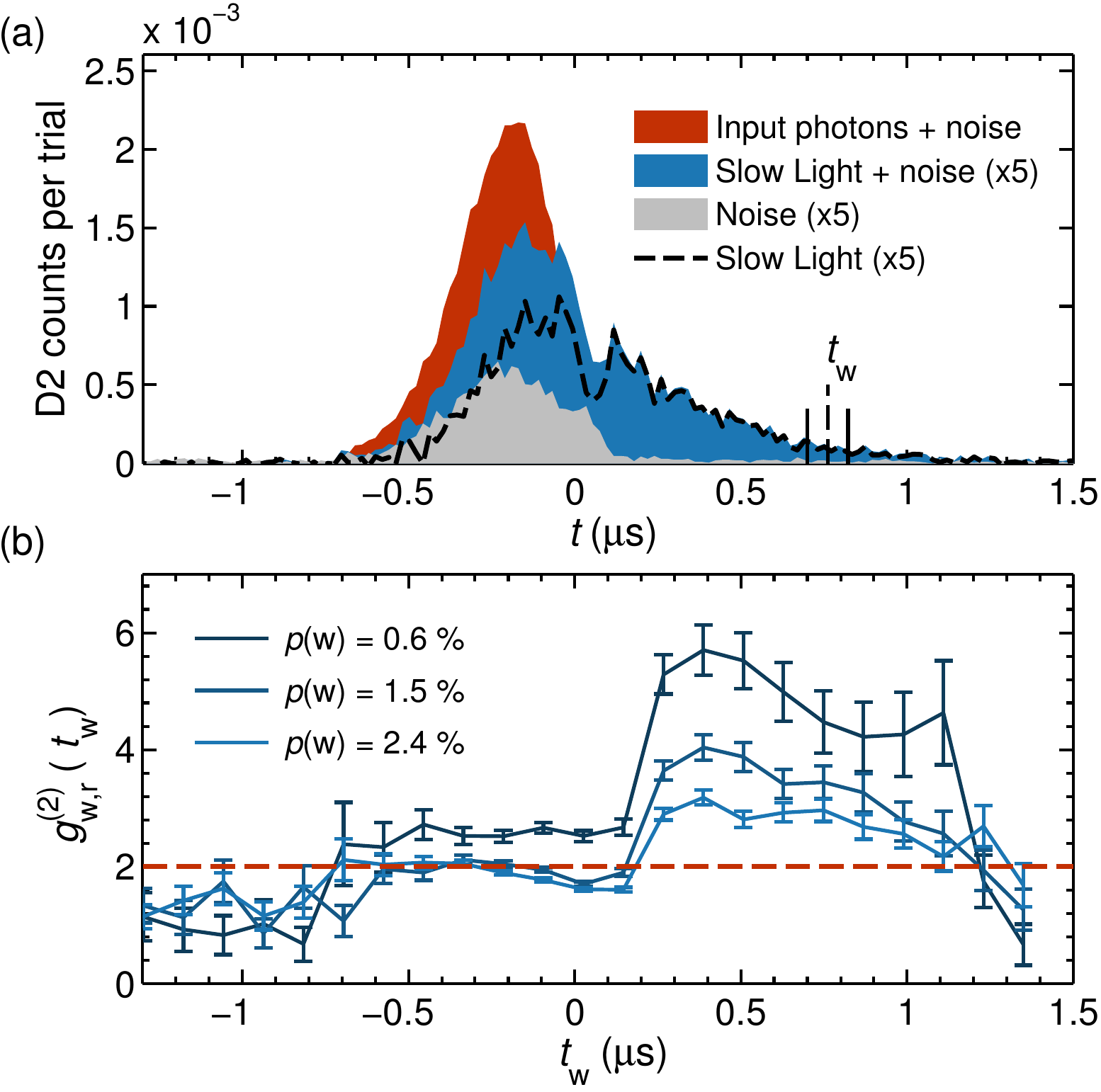}
	\caption{(a) Count rate on the SPD D2 when no atoms are loaded in site B (red), when the atoms are loaded and the pulse is slowed down (blue), when the atoms are loaded and the coupling beam is switched off, representing the unwanted noise on the $\ket{5{\rm} S_{1/2}, F=1} \leftrightarrow \ket{5{\rm} P_{3/2}, F=2}$ transition (grey). The dashed line is the noise subtracted slowed read pulse. The dip in the slowed pulse that is observable at $t \sim 0 \, \mu$s results from the fast switch-off of the trailing edge of the input photon (see \cite{Wei2009,Zhang2011}). The two solid vertical lines show the $\sim 123 \, \rm ns$ time window gate centred at $t_{\rm w}$ (marked by the dashed vertical line) that we use to measure the cross-correlation function in (b). (b) Cross-correlation function as a function of $t_{\rm w}$. The horizontal line represents the classical bound $g^{(2)}_{\rm w,r} = 2$.}
	\label{g2eit:fig}
\end{figure}
\begin{figure}[h]
	\centering
	\includegraphics[width=9cm]{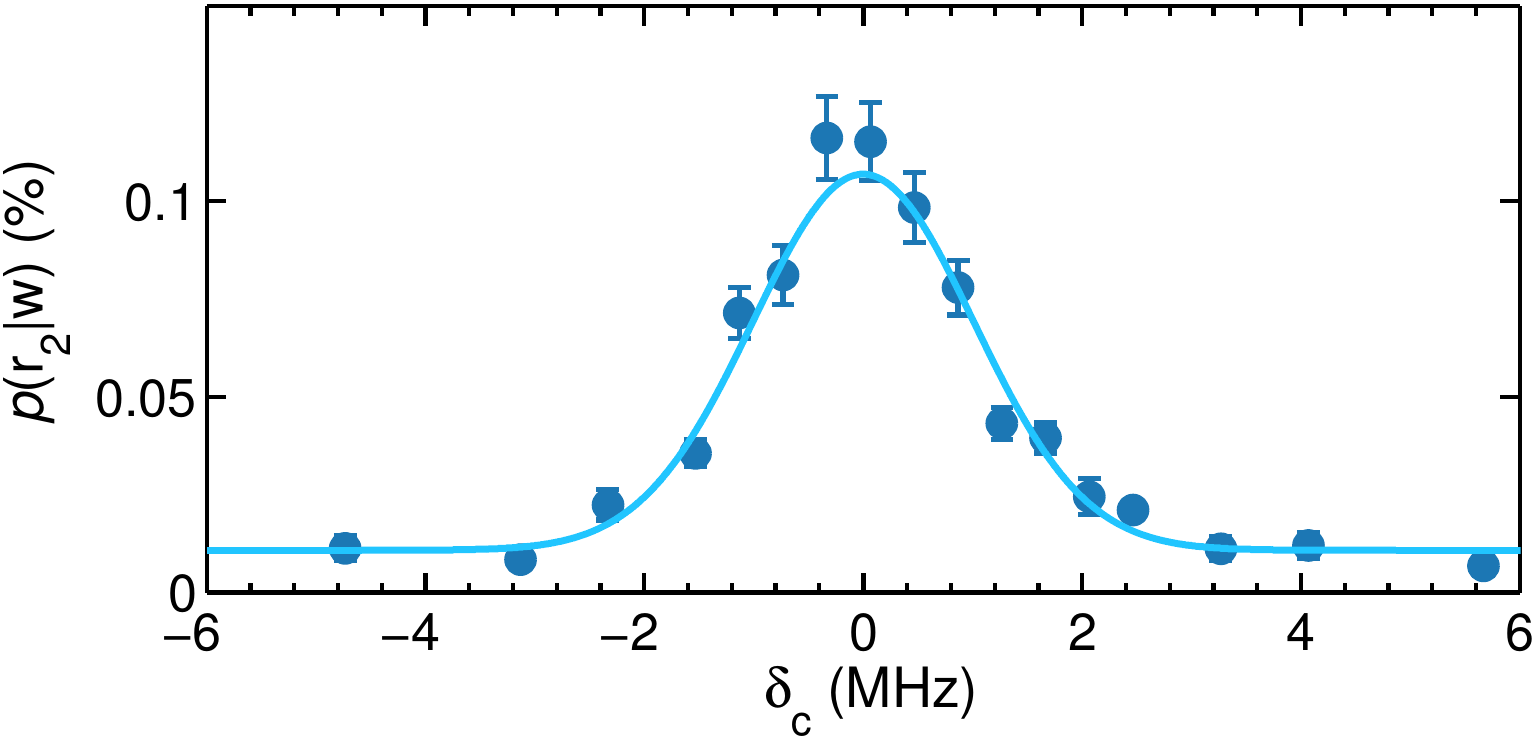}
	\caption{Coincidence detection probability $p(r_2|w)$ as a function of the the coupling beam detuning $\delta_c$ after storing the read photon for $t_{\rm R} = 500$ ns. The fit with the function $A e^{-\delta_c^2/2\sigma^2}$ (solid line) gives a width $\sigma = 1.01 \pm 0.04$ ($\rm FWHM = 2.38 \pm 0.09\,\rm MHz$). The error bars are the Poissonian error of the photon counting statistics.}
	\label{bandwith:fig}
\end{figure}

\end{document}